Title

# Profiling Of Volatiles In Tissues Of *Salacia Reticulata* Wight. With Anti-Diabetic Potential Using GC-MS And Molecular Docking

Authors

Anurabh Charkravarty[1], Gnanam Ramasamy[2]

1. Department of Plant Biotechnology, Centre for Plant Molecular Biology and Biotechnology, Tamil Nadu Agricultural University, Coimbatore-03
   Email: anurabh.chakravarty@gmail.com
   ORCID: https://orcid.org/0000-0001-5781-9381
2. Department of Plant Molecular Biology and Bioinformatics, Centre for Plant Molecular Biology and Biotechnology, Tamil Nadu Agricultural University, Coimbatore-03
   Email: gnanam.r@tnau.ac.in
   ORCID: http://orcid.org/0000-0001-9894-2140
   SCOPUS ID: https://www.scopus.com/authid/detail.uri?authorId=8904136400

1. Abstract:

Type 2 diabetes mellitus is a global pandemic, it is a chronic, progressive and an incompletely understood metabolic condition. The disease is characterized by higher levels of sugar in blood caused either due to insufficient production of insulin or because of insulin resistance. Major drugs used for the treatment of the condition are fraught with side effects. Hence, it is becoming an obligation to gaze at alternative agents showing marginal adverse effects. An important source of such agents are the medicinal plants. Several plants have been positively identified to show anti-diabetic effects. The species *Salacia reticulata* Wight., belonging to the family Celastraceae which is found in the forests of Southern India is one such promising plant to tackle type 2 diabetics. In this study, numerous volatile compounds were identified from various tissues through GC-MS analysis. Among these the compounds possessed suitable ADMET properties, and high binding affinities were compared with two approved α-glucosidase inhibitors, Acarbose and Miglitol. The analysis indicated that the compounds with PubChem IDs, CID-240051 and CID-533471 exhibited potential as inhibitors of Human Maltase-Glucoamylase enzyme.

Introduction:

*Salacia reticulata* Wight., is a large woody climbing shrub found in South Indian states ranging from Southern Odisha to Kerala as well as Sri Lanka. It is commonly called 'Ponkoranti' in Tamil, 'Ekanayakam' in Malyalam, 'Anukudu cettu' in Telegu, 'Saptaragni' in Sanskrit and 'Kothala himbutu' in Sinhala. The plant has been used in traditional medicine to treat various ailments in India. The species was formerly placed in the family Hippocrateaceae but has now been moved to the family Celastraceae which is colloquially known as the staff-vine or bittersweet family. Many bioactive compounds have been reported for the genus *Salacia* (Arunakumara and Subasinghe 2011). Several phenolics, flavonoids, and antioxidants from *Salacia chinensis* L. and communicated that when the field grown plants were sprayed with aqueous extracts of biotic and abiotic elicitors, further more eight new compounds were identified in GC-MS (Ghadage et al. 2017). Though the chemical composition of different parts of the plant has been extensively studied, yet the role of volatile compounds found in the plant has not been studied so far.

Extracts of *Salacia reticulata* Wight., roots has been utilized to treat inflammations, asthma, amenorrhea and dysmenorrhea (Tissera and Thabrew 2001). The roots of this species are acrid, bitter, thermogenic, urinary, astringent, anodyne, anti-inflammatory (Nadkarni 1993). The roots and stem of *Salacia reticulata* Wight., have been widely used in treating diabetes and obesity (Im et al. 2008; Li et al. 2008), gonorrhea and rheumatism, skin diseases (Matsuda et al. 2002; Im et al. 2008) and hemorrhoids' (Nadkarni 1993). It was also shown that the water extracts of leaves of *Salacia reticulata* Wight., could be beneficial for the prevention of diabetes and obesity as its multiple effects such as the ability to increase the plasma insulin level and lower the lipid peroxide level of the kidney (Yoshino et al. 2009). Mangiferin, kotalanol and salacinol were reported to be the key anti-diabetic principles of *Salacia reticulata* Wight., established through pharmacological studies (Yoshikawa et al. 1997, 1998, 2001). Also a thiocyclitol compound has been isolated from *Salacia reticulata* Wight., which possesses hypoglycaemic properties (Hiromi 2008). In addition, numerous other constituents (dulcitol, kotalagenin, maytenfolic acid, soiguesterin and tannins) have also been isolated from this plant, but the exact effects of these constituent on the human physiology is not well understood (Arunakumara and Subasinghe 2011).

The inhibition of α-glucosidase has been identified as the key activity responsible for the hypoglycaemic of *Salacia reticulata* Wight. The effect has been studied using bioassay guided segregation of the various active fractions (Matsuda et al. 2005). The intestinal enzymes, α-glucosidase and α-amylase are responsible for the breakdown of starches, dextrins, maltose and sucrose into more easily absorbable monosaccharides in the small intestine. The inhibition of these enzymes leads to delayed absorption of glucose and hence attenuates the postprandial glucose surges in diabetic individuals. This mechanism is currently being exploited by α-glucosidase inhibitors like the drugs Acarbose and Voglibose (Derosa et al. 2012). Hence, extracts of *Salacia reticulata* Wight., reduces the postprandial hyperglycaemia and hyperinsulinemia by slowing the catabolism of poly and oligo saccharides. The methanolic as well as aqueous methanolic extracts from *Salacia reticulata* Wight., has been shown to significantly reduce glucose levels in maltose and sucrose loaded rats. While, this inhibition was not observed in glucose loaded rats, confirming the inhibitory effect of *Salacia reticulata* Wight., on the intestinal brush border enzymes (Matsuda et al. 2005).

It has been observed that aqueous extract of *Salacia reticulata* Wight. in a dose dependent manner suppresses the serum glucose level induced by sucrose, maltose and α-starch but not that induced by glucose and lactose (Kawamori and Kawahara 1998). The study also reported the effect of the *Salacia reticulata* Wight., extract on the activities of various rat jejunum derived glucosidases and α-amylase. Concluding that *Salacia reticulata* Wight., extracts strongly inhibited the action of α-glucosidase and α-amylase but not that of β-glucosidase enzyme. Similarly, another study reported the ability of *Salacia reticulata* Wight., extract to reduce the postprandial levels of glucose in sucrose and maltose loaded male Wistar rats to significantly at 0.5 and 2 hours whereas, glucose inhibition was not observed in the glucose loaded rats (Yoshikawa et al. 2002). The study attributed the glucose lowering effect to the compounds Salacinol and Kotalanol (Matsuda et al. 2005).

Another study reported the sucrase and maltase inhibitory activities of the aqueous extract of *Salacia reticulata* Wight., by using maltose and sucrose as substrates *in vitro*. Through the study it was observed that the novel thiocyclitol compound demonstrated greater α-glucosidase inhibitory effects compared to the compound Salacinol. In addition the study also analyzed the postprandial levels of glucose in maltose and sucrose loaded rats after administering them with either *Salacia reticulata* Wight., extract, Salacinol, a novel thiocyclitol compound or Voglibose. The results of the study confirmed that both maltase and sucrase activity were inhibited by all these compounds albeit the inhibitory activity observed was several times weaker than that of the drug Voglibose (Hiromi 2008).

Innumerable drugs fail to penetrate the market owing to poor pharmacokinetic properties, which leads to enormous losses to pharmaceutical companies (Fang et al. 2018). The emergence of *in silico* tools as sophisticated methods for drug discovery, has been applied to screen potential drugs from phytochemicals found in various medicinal plants (Sliwoski et al. 2014). Computational prediction models, also play a crucial role in the process of selection of methodologies guiding pharmaceutical research, and hence have also been utilized in *in silico* prediction of pharmacological, pharmacokinetic and toxicological properties of compounds (Loza-Mejía et al. 2018). Presently, molecular docking serves as an efficient and economical method for designing and testing of drugs. The technique generates information pertaining to the drug receptor interactions that are helps predict the binding mode of the drug candidates to their target (Lee and Kim 2019). Furthermore, this methodology facilitates a systemic study by introducing a molecule to the binding spot of the target in a non-covalent fashion, enabling accurate binding at the active site of the ligand (Bharathi et al. 2014). Therefore, the present study focuses on identification of bioactive compounds from methanol extracts from *Salacia reticulata* Wight., root, root bark, stem and leaf by GC–MS analysis. The study also evaluated the efficacy of the various identified volatiles of *Salacia reticulata* Wight., for their antidiabetic activity through, *in silico* molecular docking analysis.

## Methods:

### Collection and processing of plant material

Plant materials from root, root bark, stem, and leaf of *Salacia reticulata* Wight., were collected from the Tamil Nadu Agricultural University, Botanical Garden at Coimbatore (11°01′05.3′′N; 76°55′58.1″E) Tamil Nadu, India. The samples were thoroughly dried and ground prior to extraction.

### Continuous shaking extraction

One gram powder of each sample (root, root bark, stem and leaf) was subjected to continuous shaking on a shaker (C1 Platform Shaker, New Brunswick Scientific, Edison, NJ, USA) for extraction using 50 ml of methanol and placed at a room temperature (25 ± 2 °C) at 110 rpm for 24 hours. The extract was then filtered using a Whatman No. 1 filter paper and was concentrated to 20 ml.

### GC-MS analysis of phytoconstituents

Volatiles from each sample (root, root bark, stem and leaf) were identified by GC/MS using Perkin Elmer Clarus SQ 8C equipped with an Elite-5MS (100% Dimethyl poly siloxane), 30 x 0.25 mm x

0.25 µm column. A turbo mass-gold Perkin-Elmer detector was used. The flow rate for the carrier gas was set at 1 ml per minutes, split 10:1, and injected volumes were 1µl. The column temperature was initially raised to 110°C at the rate of 10°C/minutes, following which the temperature was further raised to 280°C at the rate of 5°C /min for 9 minutes. The injector temperature was set at 250°C and this temperature was held constant for a duration of 36 minutes. The electron impact energy was 70eV, the source temperature was set at 200°C. Electron impact (EI) mass scan (m/z) was recorded in the range of 45-450 amu. Unknown components were compared with the known mass spectra of compounds provided by National Institute of Standards and Technology (NIST).

### Preparation of ligand library

The structures of the identified compounds were downloaded from PubChem. The structures of two alpha-glucosidase inhibitors have also been included to enable comparison.

### Preparation of target

The structure 3TOP (Human Maltase-Glucoamylase) was chosen as it was bound to the drug Acarbose which would permit validation of the screening strategy by redocking (Ren et al. 2011). The binding pocket used for the docking study was based on the interacting residues as described in Table 1.

### ADMET properties

The ADMET properties were predicted using DruLiTo software (http://www.niper.gov.in/pi_dev_tools/DruLiToWeb/DruLiTo_index.html). The libraries were filtered stringently based on multiple filters namely, Lipinski's rule, BBB likeness, unweighted QED and weighted QED. Only compounds passing all the filters were selected.

### Virtual Screening

The compounds possessing suitable ADMET properties were then screened against Human Maltase-Glucoamylase to ascertain their binding affinity using MTiOpenScreen web-server (https://mobyle.rpbs.univ-paris-diderot.fr/cgi-bin/portal.py#forms::MTiOpenScreen) (Labbé et al. 2015). The four hits with highest binding affinity were docked again using PyRx 8.0 with exhaustiveness set at 12. The interaction between the ligands and receptor were analysed using PLIP web-server (https://plip- tool.biotec.tu-dresden.de/plip-web/plip/index) (Salentin et al. 2015).

Result & Discussion:

GC-MS analysis

The volatile compounds identified from the aerial tissues viz. leaf and stem tissues by GC-MS have been tabulated in Table 2 and Table 3 while for the below-ground tissues viz., root and root bark in Table 4 and Table 5 respectively along with their retention times, IUPAC names and Area Percentage. The chromatograms generated have been depicted in Fig. 1.

A total of 160 peaks were analysed resulting in the identification of 99 compounds. The number of unique compounds for leaf, stem, root and root bark are 22, 15, 16 and 21 respectively. The distribution of the volatiles in the analysed tissue has been represented in the form of a Venn diagram in Fig. 2. The similarity profile of the tissues studied has been represented in the form of a pairwise similarity matrix based on the number of non-distinct volatiles present among the individual pairs in Fig. 3.

Virtual screening

The compounds which have suitable ADMET properties based on the filtering criterion and the highest binding affinities are 6-Hydroxy-7-isopropyl-1,4a-dimethyl-1,2,3,4,4a,9,10,10a-octahydro- 1-phenanthrenemethanol,(1à, 4aá, 10a.alpha)-(CID – 240051),1-Oxaspiro[4.4]non-8-ene-4,7- dione,9-hydroxy-6-(3-methyl-2-butenyl)-2-(1-methylethyl)-8-(3-methyl-1-oxobutyl)- (CID – 533471) 2,4-Di-tert-butylphenol (CID – 7311), 16-Nitrobicyclo[10.4.0]hexadecan-1-ol-13-one (CID- 544211), their respective binding affinities are -11.01, -10.28, -8.17, -6.11 kcal/mol. The redocked Acarbose molecule as compared to the native Acarbose bound to Human Maltase-Glucoamylase (3TOP) can be viewed in Fig. 4. The four compounds are with their binding affinity and plant tissues are furnished in Table 6. The binding affinities of the approved drugs Acarbose and Miglitol are -10.93 and -7.29 kcal/mol. The interactions between CID – 240051 and Human Maltase-Glucoamylase (3TOP) have been elucidated in Table 3. CID – 240051 and CID – 533471 can be seen bound to Human Maltase-Glucoamylase (3TOP) in Fig. 5. The compound CID – 240051 forms hydrogen bonds *via* the residue TYR1251A while the compound CID – 533471 does so with the residue ARG1501A.

The objective of the study was to identify novel secondary metabolites from various tissues of *Salacia reticulata* Wight., and to identify their anti-diabetic potential through *in silico* analysis. The phytochemical profile for the different tissues was generated and compared. The identification of secondary metabolites was done using GC-MS. GC-MS is a robust analytical method to identify volatiles. GC-MS has been used to generate metabolites in several plant species such as *Ximenia americana*, *Nauclea latifolia*, *Rhazya stricta*, *Androsace foliosa* to name a few (Shettar et al. 2017; Mahmood et al. 2020; Mgbeje and Abu 2020; Zaheer et al. 2021). In the current study, among the ninety-nine different compounds present identified in the tissues, the compounds possessing optimal ADMET properties were chosen for molecular docking against the target, human intestinal brush-border maltase-glucoamylase enzyme to identify the compounds with anti-diabetic potential. Earlier studies have studies have shown that phytochemicals show inhibition of human maltase-glucoamylase enzyme through molecular docking analysis. The triterpenes α-amyrine, β- amyrine, ursolic acid, oleanolic acid, betulinic acid found in *Pelliciera rhizophorae* show greater inhibition of the enzyme as compared to the drug Acarbose. In another study the compounds salacinol, kotalanol, and de-O-sulfonated kotalanol found in the species *Salacia reticulata* Wight., were shown to have excellent affinity for human maltase glucoamylase (Sim et al. 2010).

From our study the terpenoids with PubChem IDs CID – 240051 and CID – 533471 show true potential as inhibitors for the enzyme. Compounds having α-glucosidase inhibitory activity are ubiquitous in medicinal plants and commercially available glucosidase inhibitors involves tedious synthesis process and are associated with gastrointestinal side-effects. The compounds identified, 1,2,3,4,4a,9,10,10a-Octahydro-6-hydroxy-7-isopropyl-1,4a-dimethyl-1-phenanthrenemethanol (CID-240051) and 1-Oxaspiro[4.4]non-8-ene-4,7-dione, 9-hydroxy-6-(3-methyl-2-butenyl)-2-(1- methylethyl)-8-(3-methyl-1-oxobutyl)- (CID-533471) show potential as inhibitors of alpha- glucosidase enzyme. The identified natural compounds could have tremendous clinical potential compared to synthetic oral hypoglycemic drugs after systematic evaluation and validation.

Table 1: Active residues of α-glucosidase.

| PDB ID | Active Residues | Source |
|---|---|---|
| 3TOP | TRP1418A, TYR1251A, TRP1355A, TRP1369A, TRP1523A | (Ren et al. 2011) |

Table 2: Volatiles identified in leaf.

| Name of Compound | RT | Area % | Name of Compound | RT | Area % |
|---|---|---|---|---|---|
| Dodecane | 5.764 | 0.366 | Octadecanoic acid | 22.336 | 17.372 |
| Benzene, 1,3-bis(1,1-dimethylethyl)- | 6.44 | 0.645 | 2-Nonadecanone 2,4-dinitrophenylhydrazine | 23.296 | 0.421 |
| Tetradecane | 8.481 | 0.435 | Ethanol, 2-(9,12-octadecadienyloxy)-,(Z,Z)- | 23.391 | 2.819 |
| 1,2-O-Isopropylidene-D-xylofuranose, TBDMSderivative | 9.741 | 1.637 | 2-Nonadecanone 2,4-dinitrophenylhydrazine | 23.766 | 0.437 |
| 2,4-Di-tert-butylphenol | 9.821 | 1.226 | Oxiranepentanoic acid, 3-undecyl-, methylester, trans- | 23.801 | 0.557 |
| Undecanoic acid, 10-methyl-, methyl ester | 10.066 | 0.504 | 1-Oxaspiro[4.4]non-8-ene-4,7-dione, 9-hydroxy-6-(3-methyl-2-butenyl)-2-(1-methylethyl)-8-(3-methyl-1-oxobutyl)- | 23.931 | 0.597 |
| Octadecane, 3-ethyl-5-(2-ethylbutyl)- | 10.201 | 0.306 | Cyclohexane, 1,1'-dodecylidenebis[4-methyl- | 24.067 | 0.631 |
| Pentadecanoic acid | 10.711 | 0.254 | 5H-Cyclopropa[3,4]benz[1,2-e]azulen-5-one, 3,9,9a-tris(acetyloxy)-3-[(acetyloxy)methyl]-2-chloro-1,1a,1b,2,3,4,4a,7a,7b,8,9,9a-dodecahydro- 4a,7b-dihydroxy-1,1,6,8-tetramethyl-, [1aR-(1aà,1bá,2à,3á,4aá,7aà,7bà,8à,9á,9aà)]- | 24.197 | 0.511 |
| 1,4-Dioxane-2,5-diol,2TBDMS derivative | 12.382 | 0.686 | Prost-13-en-1-oic acid, 9,11,15-trihydroxy-6-oxo-, methyl ester, (9à,11à,13E,15S)- | 24.337 | 0.253 |
| à-D-Glucofuranose, 6-O-(trimethylsilyl)-, cyclic 1,2:3,5-bis(butylboronate) | 12.562 | 0.306 | Cyclopropanebutyric acid, 2-[(2-nonylcyclopropyl)methyl]-, methyl ester | 24.482 | 0.448 |
| Heptadecane, 9-hexyl- | 13.377 | 0.302 | Ursodeoxycholic acid | 24.572 | 0.59 |
| Neophytadiene | 16.199 | 0.279 | Glycyl-L-histidyl-L-lysine acetate | 24.652 | 0.443 |
| Stearyltrimethylammoniumchloride | 17.604 | 0.7 | 2,6,9,12,16-Pentamethylheptadeca-2,6,11,15-tetraene-9-carboxylic acid | 24.802 | 0.578 |
| Pentadecanoic acid, 14-methyl-, methyl ester | 17.959 | 0.303 | 17-Pentatriacontene | 25.092 | 0.304 |
| 1,2-Benzenedicarboxylic acid, butyl octyl ester | 18.399 | 0.27 | 1-Diphenylethenylsilyloxydodec-9-yn | 25.332 | 0.375 |
| n-Hexadecanoic acid | 18.654 | 5.835 | Squalene | 27.578 | 2.751 |
| Ethyl iso-allocholate | 19.26 | 0.282 | Glucobrassicin | 27.863 | 0.353 |
| 16-Octadecenoic acid, methylester | 21.175 | 0.282 | Hexadecanoic acid, 1-(hydroxymethyl)-1,2-ethanediyl ester | 28.053 | 0.774 |
| 9,12-Octadecadienoic acid(Z,Z)- | 21.756 | 4.147 | Diisooctyl phthalate | 28.283 | 1.254 |
| 9-Octadecenoic acid, (E)- | 21.906 | 30.426 | Cholest-22-ene-21-ol, 3,5-dehydro-6-methoxy-, pivalate | 29.234 | 0.25 |

Table 3: Volatiles identified in stem.

| Name of Compound | RT | Area % | Name of Compound | RT | Area% |
|---|---|---|---|---|---|
| Undecane | 4.319 | 0.421 | 2-Ethylbutyric acid, eicosyl ester | 13.372 | 2.311 |
| Dodecane | 5.769 | 1.141 | Epirubicin | 13.633 | 0.47 |
| Benzene, 1,3-bis(1,1-dimethylethyl)- | 6.435 | 1.174 | Hexadecanoic acid, methyl ester | 17.954 | 0.364 |
| Dodecane, 2,6,11-trimethyl- | 6.8 | 0.376 | n-Hexadecanoic acid | 18.655 | 4.187 |
| Tetradecane | 8.486 | 1.592 | 9,12-Octadecadienoic acid (Z,Z)- | 21.751 | 3.069 |
| 1,2-Propanediol, 3-(octadecyloxy)-, diacetate | 9.261 | 0.479 | 9-Octadecenoic acid, (E)- | 21.901 | 21.918 |
| Octadecane, 3-ethyl-5-(2-ethylbutyl)- | 9.516 | 0.456 | Octadecanoic acid | 22.336 | 9.027 |
| Dodecane, 2,7,10-trimethyl- | 9.566 | 0.566 | 1-Hexadecanol, 2-methyl- | 22.911 | 1.195 |
| 1,2-O-Isopropylidene-D-xylofuranose, TBDMS derivative | 9.731 | 3.633 | Dasycarpidan-1-methanol, acetate (ester) | 23.001 | 1.95 |
| 2,4-Di-tert-butylphenol | 9.821 | 1.86 | Oleic Acid | 23.406 | 1.208 |
| à-D-Galactopyranose, 6-O-(trimethylsilyl)-, cyclic1,2:3,4-bis(methylboronate) | 9.991 | 0.354 | 16-Nitrobicyclo[10.4.0]hexadecan-1-ol-13- one | 23.511 | 0.504 |
| 9-Bromononanoic acid,methyl(ester) | 10.066 | 0.736 | Dasycarpidan-1-methanol, acetate (ester) | 23.822 | 0.467 |
| Hexadecane, 5-butyl- | 10.126 | 0.369 | 1b,4a-Epoxy-2H-cyclopenta[3,4]cyclopropa[8,9]cycloundec[1, 2-b]oxiren-5(1aH)-one, 2,7,9,10-tetrakis(acetyloxy)decahydro-3,6,8,8, 10a-pentamethyl- | 23.887 | 0.478 |
| Eicosane, 2-methyl- | 10.196 | 0.501 | 7-Methyl-Z-tetradecen-1-ol acetate | 24.082 | 0.436 |
| Hexadecane | 11.397 | 0.65 | Dasycarpidan-1-methanol, acetate (ester) | 24.477 | 0.424 |
| Pentitol, 1,3-didesoxy-tris-O-(trimethylsilyl)- | 11.842 | 1.25 | (5á,13á) Androst-8-en-3-one, 17-19-diacetoxy-4,4-dimethyl- | 24.817 | 0.386 |
| 1,4-Dioxane-2,5-diol,2TBDMS derivative | 12.037 | 0.969 | Octadecane, 3-ethyl-5-(2-ethylbutyl)- | 25.092 | 0.591 |
| 1,4-Dioxane-2,5-diol,2TBDMS derivative | 12.367 | 3.042 | Octadecane, 3-ethyl-5-(2-ethylbutyl)- | 25.772 | 0.385 |
| 1,4-Dioxane-2,5-diol,2TBDMS derivative | 12.547 | 1.389 | Squalene | 27.583 | 2.539 |
| 2-Oxiraneethanol, 2-t-butyldimethysilyloxymethyl-acetate | 13.282 | 2.84 | Diisooctyl phthalate | 28.293 | 0.544 |

Table 4: Volatiles identified in root.

| **Name of Compound** | **RT** | **Area %** | **Name of Compound** | **RT** | **Area%** |
|---|---|---|---|---|---|
| Undecane, 4,7-dimethyl- | 3.639 | 0.692 | 1,2-Benzenedicarboxylic acid, butyl 8-methylnonyl ester | 18.394 | 0.707 |
| Undacane | 4.289 | 0.891 | n-Hexadecanoic acid | 18.62 | 2.286 |
| Dodecane | 5.739 | 2.231 | Hexadecanoic acid, ethyl ester | 19.255 | 0.542 |
| Benzene, 1,3-bis(1,1-dimethylethyl)- | 6.41 | 2.428 | Ethanol, 2-(9,12-octadecadienyloxy)-, (Z,Z)- | 21.025 | 0.613 |
| Hexadecane | 6.775 | 0.917 | 9-Octadecenoic acid (Z)-, methyl ester | 21.166 | 0.929 |
| Tetradecane | 8.461 | 2.793 | 9,12-Octadecadienoic acid (Z,Z)- | 21.731 | 1.765 |
| Undecane, 3,7-dimethyl- | 7.41 | 0.556 | 9-Octadecenoic acid, (E)- | 21.856 | 10.67 |
| Dodecanal | 8.586 | 0.643 | Octadecane, 3-ethyl-5-(2-ethylbutyl)- | 22.091 | 1.444 |
| Dodecane, 5,8-diethyl- | 9.166 | 0.524 | 12-Methyl-E,E-2,13-octadecadien-1-ol | 22.251 | 1.184 |
| Tetradecane, 2,6,10-trimethyl- | 9.236 | 1.249 | Octadecanoic acid | 22.321 | 2.752 |
| Octadecane, 3-ethyl-5-(2-ethylbutyl)- | 9.316 | 0.612 | 1-Hexadecanol, 2-methyl- | 22.886 | 0.838 |
| Dodecane, 2,6,11-trimethyl- | 9.546 | 1.298 | Octadecane, 3-ethyl-5-(2-ethylbutyl)- | 25.077 | 0.708 |
| 2,4-Di-tert- butylphenol | 9.801 | 0.817 | Ethanol, 2-(9-octadecenyloxy)-, (Z)- | 26.908 | 4.076 |
| Dodecane, 1-iodo- | 10.171 | 0.774 | 17-Pentatriacontene | 27.373 | 0.739 |
| Hexadecane | 11.372 | 0.94 | 2,6,10,14,18-Pentamethyl-2,6,10,14,18-eicosapentaene | 27.583 | 3.738 |
| Pentadecane, 2,6,10,14-tetramethyl- | 13.347 | 0.912 | Retinoyl-á-glucuronide 6',3'- lactone | 28.128 | 0.668 |
| Heptadecane, 9-hexyl- | 13.422 | 0.766 | Diisooctyl phthalate | 28.288 | 2.132 |
| Heptadecane, 2,6,10,15-tetramethyl- | 14.243 | 0.794 | Octadecane, 3-ethyl-5-(2- ethylbutyl)- | 28.393 | 1.025 |
| 1-Nonadecene | 15.318 | 0.997 | 6-Hydroxy-7-isopropyl-1,4a- dimethyl-1,2,3,4,4a,9,10,10a- octahydro-1- phenanthrenemethanol, (1à, 4aá,10a.alpha)- | 29.314 | 0.836 |
| Hexadecanoic acid,methyl ester | 17.944 | 0.887 | Phenol, 2,3,4-trimethoxy-5-propionyloxy-6-propionyl- | 29.404 | 1.093 |

Table 5: Volatiles identified in root bark.

| Name of Compound | RT | Area % | Name of Compound | RT | Area % |
|---|---|---|---|---|---|
| Dodecane | 5.764 | 0.785 | Ethanol, 2-(9-octadecenyloxy)-, (Z)- | 23.422 | 2.563 |
| Benzene, 1,3-bis(1,1-dimethylethyl)- | 6.435 | 1.392 | Oxiraneoctanoic acid, 3-octyl-, cis- | 23.612 | 0.933 |
| Dodecane, 2,6,10-trimethyl- | 6.8 | 0.321 | Glycidyl oleate | 23.812 | 1.801 |
| Tetradecane | 8.486 | 1.154 | Dodecyl cis-9,10-epoxyoctadecanoate | 24.112 | 1.548 |
| Heptacosane | 9.566 | 0.456 | 2-(16-Acetoxy-11-hydroxy-4,8,10,14- tetramethyl-3-oxohexadecahydrocyclopenta[a]phen anthren-17-ylidene)-6-methyl-hept-5- enoic acid, methyl ester | 24.337 | 0.655 |
| 2,4-Di-tert-butylphenol | 9.826 | 0.411 | Glucobrassicin | 24.537 | 1.365 |
| 2-Bromo dodecane | 10.196 | 0.371 | Ferruginol | 24.787 | 1.176 |
| Hexadecane | 11.397 | 0.486 | 9,10-Anthracenedione, 1,8- dimethoxy- | 24.887 | 0.477 |
| Tetradecane, 1-iodo- | 13.383 | 0.355 | Octadecane, 3-ethyl-5-(2-ethylbutyl)- | 25.122 | 0.781 |
| 1-Docosene | 15.343 | 0.31 | Androst-5,7-dien-3-ol-17-one | 25.262 | 0.441 |
| Hexadecanoic acid, methyl ester | 17.959 | 0.37 | Pseduosarsasapogenin-5,20-dien | 25.362 | 0.394 |
| 1,2-Benzenedicarboxylic acid, butyl decyl ester | 18.405 | 0.447 | 17-Pentatriacontene | 25.767 | 0.438 |
| n-Hexadecanoic acid | 18.655 | 4.339 | Z-8-Octadecen-1-ol acetate | 26.913 | 0.447 |
| Cholestan-3-ol, 2-methylene-, (3á,5à)- | 21.056 | 0.32 | Squalene | 27.573 | 2.274 |
| 6-Octadecenoic acid, methyl ester, (Z)- | 21.181 | 0.57 | Hexadecanoic acid, 1- (hydroxymethyl)-1,2-ethanediyl ester | 27.863 | 0.481 |
| 9,12-Octadecadienoic acid (Z,Z)- | 21.751 | 3.561 | Retinoic acid | 28.123 | 0.399 |
| 9-Octadecenoic acid, (E)- | 21.906 | 29.141 | Di-n-octyl phthalate | 28.283 | 0.824 |
| Octadecanoic acid | 22.346 | 9.103 | 14-Octadecenal | 28.393 | 0.489 |
| Oleic Acid | 22.646 | 3.725 | 6-Hydroxy-7-isopropyl-1,4a- dimethyl-1,2,3,4,4a,9,10,10a- octahydro-1-phenanthrenemethanol, (1à, 4aá, 10a.alpha)- | 29.319 | 0.333 |
| Z-(13,14-Epoxy)tetradec-11-en-1-ol acetate | 22.901 | 4.894 | Cyclopenta[d]antrhacene-6,8,11-trione,1,2,3,3a,4,5,6,6a,7,8,11,12-dodecahydro-3-(1-methylethyl) | 29.404 | 0.539 |

Table 6: Binding affinities of identified lead molecules and approved drugs.

| S. No. | Compound Name | | PubChem ID | Binding Affinity (kcal/mol) | Plant Part |
|---|---|---|---|---|---|
| 1. | Identified Compounds | 6-Hydroxy-7-isopropyl-1,4a-dimethyl-1,2,3,4,4a,9,10,10a-octahydro-1-phenanthrenemethanol, (1à, 4aá, 10a.alpha)- | 240051 | -11.01 | Root,Root Bark |
| 2. | | 1-Oxaspiro[4.4]non-8-ene-4,7-dione, 9- hydroxy-6-(3-methyl-2-butenyl)-2-(1- methylethyl)-8-(3-methyl-1-oxobutyl)- | 533471 | -10.28 | Leaf |
| 3. | | 2,4-Di-tert-butylphenol | 7311 | -8.17 | Leaf, Stem, Root and RootBark |
| 4. | | 16-Nitrobicyclo[10.4.0]hexadecan-1-ol-13-one | 544211 | -6.11 | Stem |
| 5. | Approved Drugs | Acarbose | 41774 | -10.93 | |
| 6. | | Miglitol | 441314 | -7.29 | |

Table 7: Interacting residues.

| **Compound** | **Hydrogen Bonds** | **Hydrophobic Interactions** |
|---|---|---|
| CID-240051 | TYR1251A | ILE1280A, PHE1427A, PHE1559A, |
| CID-533471 | ARG1501A | TYR1251A, ILE1280A,PHE1427A, PHE1559A, ILE1587A |

Fig. 1

a.

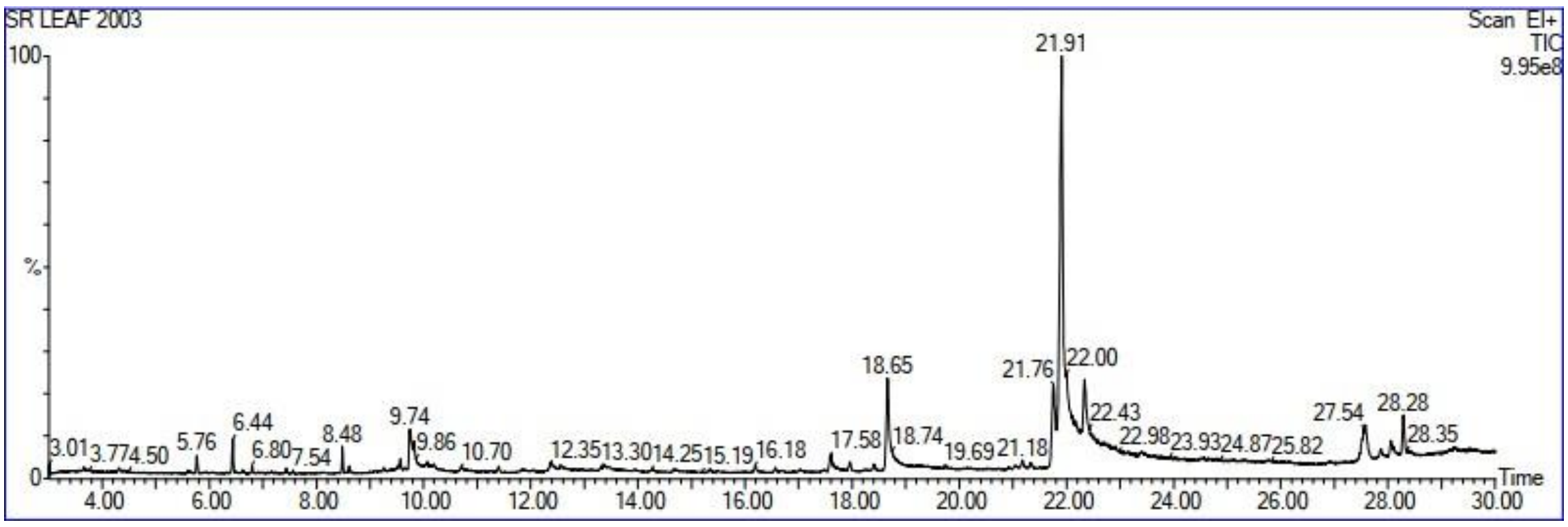
SR LEAF 2003
Scan EI+
TIC
9.95e8
100
%
0
21.91
18.65
21.76 22.00
22.43
27.54 28.28
28.35
6.44
9.74
3.01 3.77 4.50 5.76 6.80 7.54 8.48 9.86 10.70 12.35 13.30 14.25 15.19 16.18 17.58 18.74 19.69 21.18 22.98 23.93 24.87 25.82
Time
4.00 6.00 8.00 10.00 12.00 14.00 16.00 18.00 20.00 22.00 24.00 26.00 28.00 30.00

b.

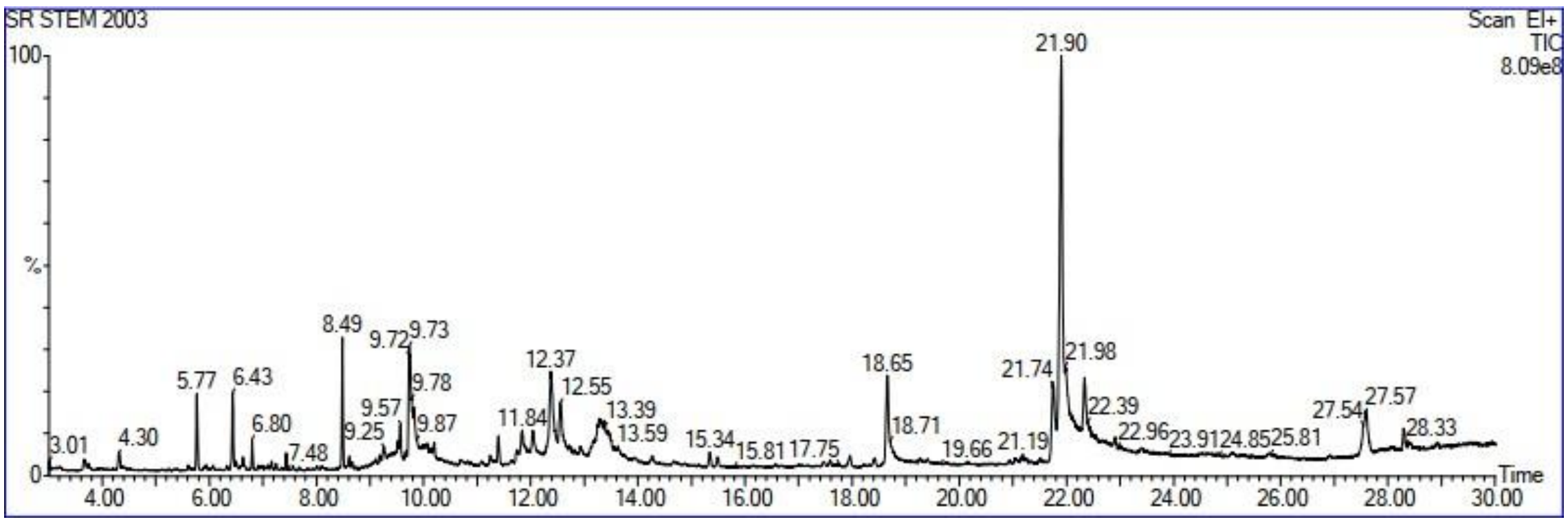
SR STEM 2003
Scan EI+
TIC
8.09e8
100
%
0
21.90
8.49 9.72 9.73
12.37
18.65
21.74 21.98
5.77 6.43
9.78
12.55
13.39
22.39
27.54 27.57
3.01 4.30 6.80 7.48 9.25 9.57 9.87 11.84 13.59 15.34 15.81 17.75 18.71 19.66 21.19 22.96 23.91 24.85 25.81 28.33
Time
4.00 6.00 8.00 10.00 12.00 14.00 16.00 18.00 20.00 22.00 24.00 26.00 28.00 30.00

c.

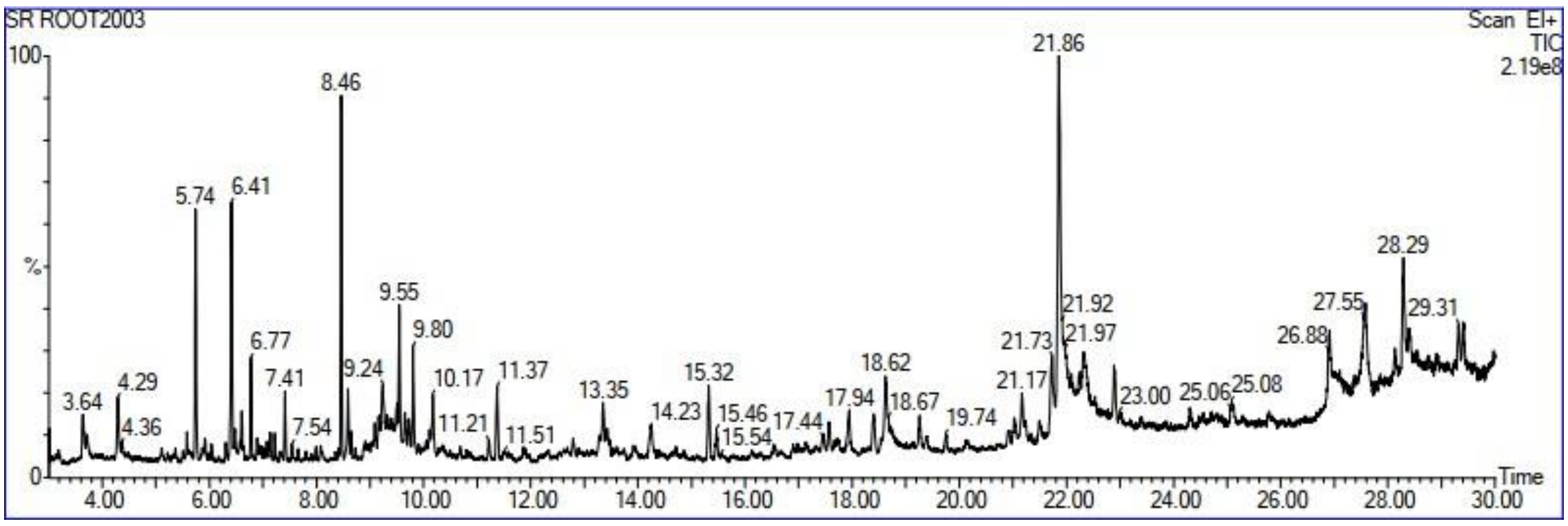
SR ROOT2003
Scan EI+
TIC
2.19e8
100
%
0
21.86
8.46
5.74 6.41
28.29
9.55
21.92
27.55 29.31
9.80
21.97
26.88
6.77 7.41 9.24 10.17 11.37 21.73
3.64 4.29 4.36 7.54 11.21 11.51 13.35 14.23 15.32 15.46 15.54 17.44 17.94 18.62 18.67 19.74 21.17 23.00 25.06 25.08
Time
4.00 6.00 8.00 10.00 12.00 14.00 16.00 18.00 20.00 22.00 24.00 26.00 28.00 30.00

d.

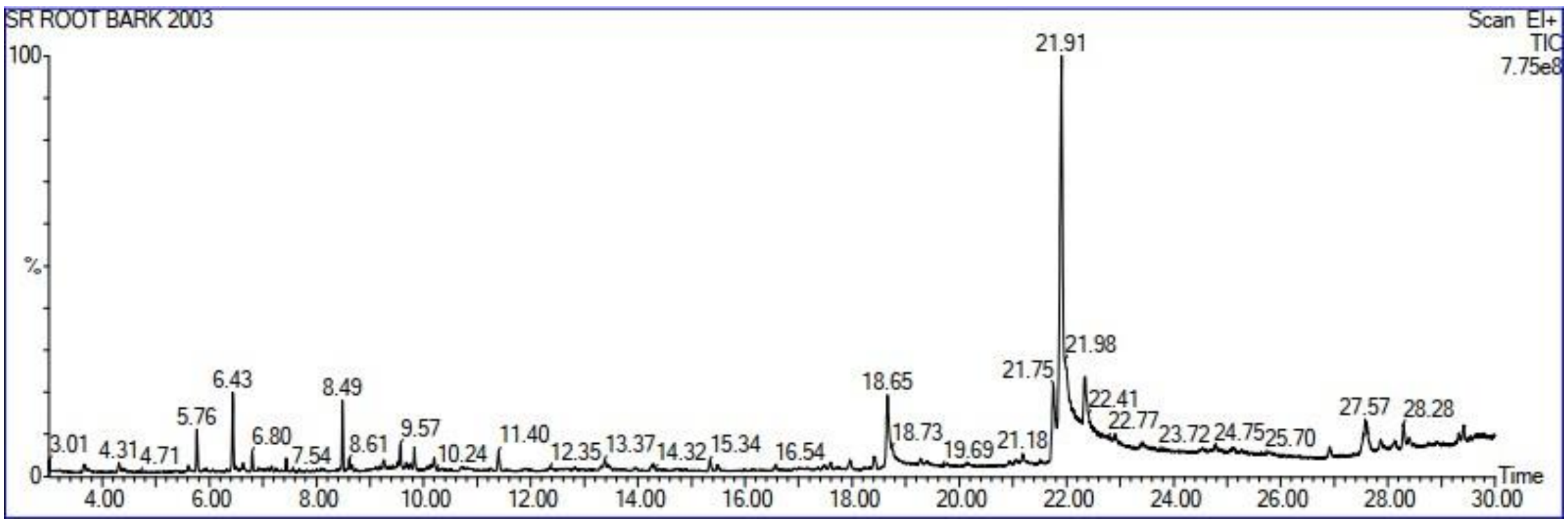
SR ROOT BARK 2003
Scan EI+
TIC
7.75e8
100
%
0
21.91
21.98
21.75
18.65
6.43 8.49
22.41
5.76
27.57 28.28
3.01 4.31 4.71 6.80 7.54 8.61 9.57 10.24 11.40 12.35 13.37 14.32 15.34 16.54 18.73 19.69 21.18 22.77 23.72 24.75 25.70
Time
4.00 6.00 8.00 10.00 12.00 14.00 16.00 18.00 20.00 22.00 24.00 26.00 28.00 30.00

Fig. 2

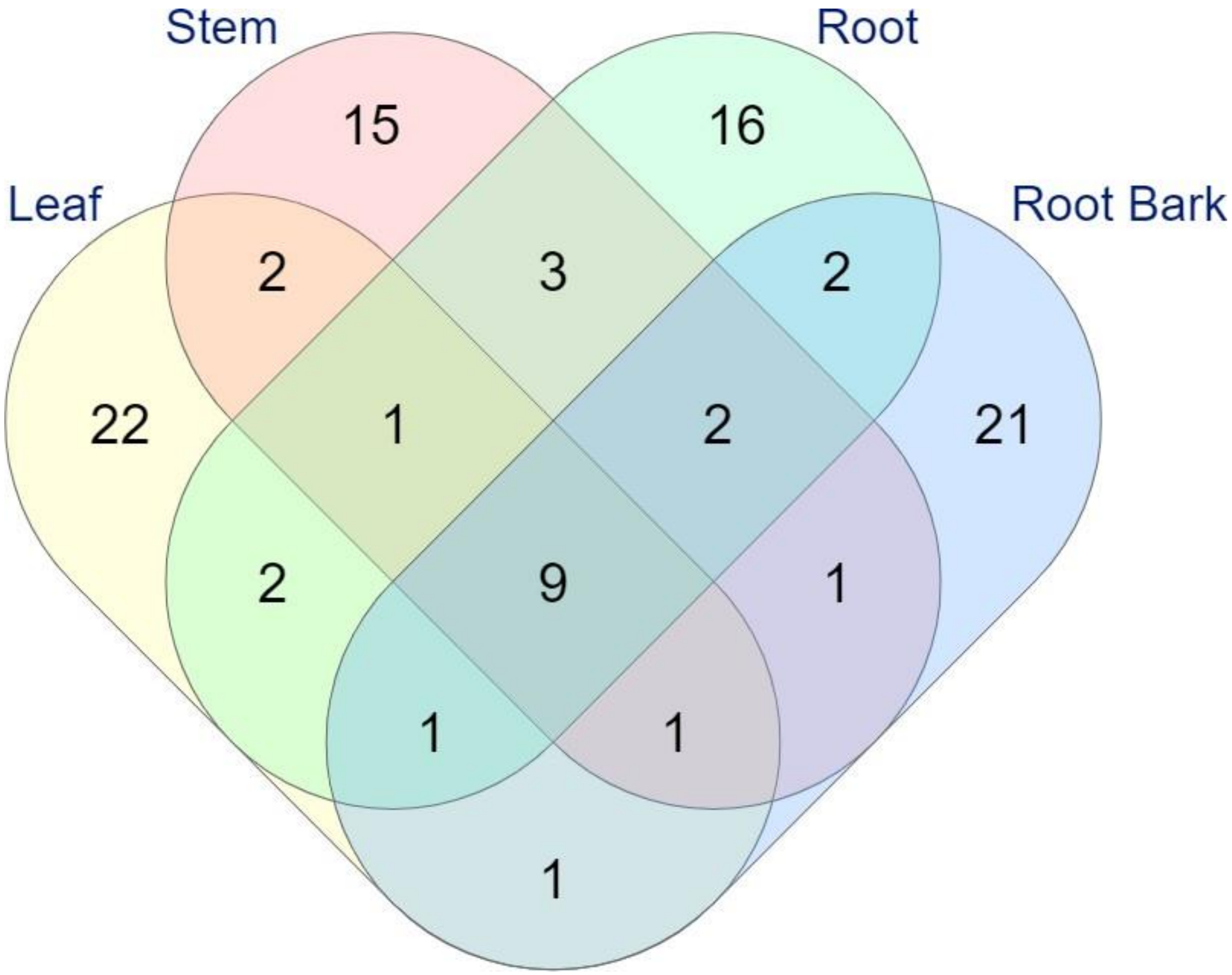

Fig. 3

| | Leaf (39) | Stem (34) | Root (36) | Root Bark (38) |
|---|---|---|---|---|
| Leaf (39) | | 13 | 13 | 12 |
| Stem (34) | 13 | | 15 | 13 |
| Root (36) | 13 | 15 | | 14 |
| Root Bark (38) | 12 | 13 | 14 | |

Fig. 4

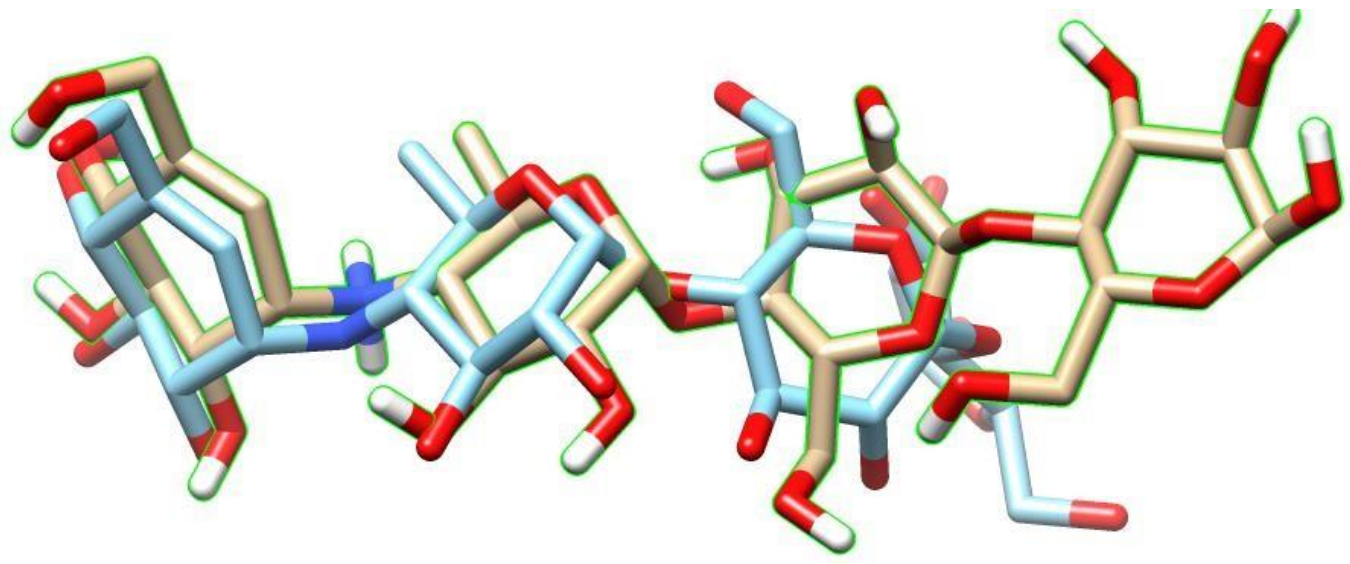

Fig. 5

a.

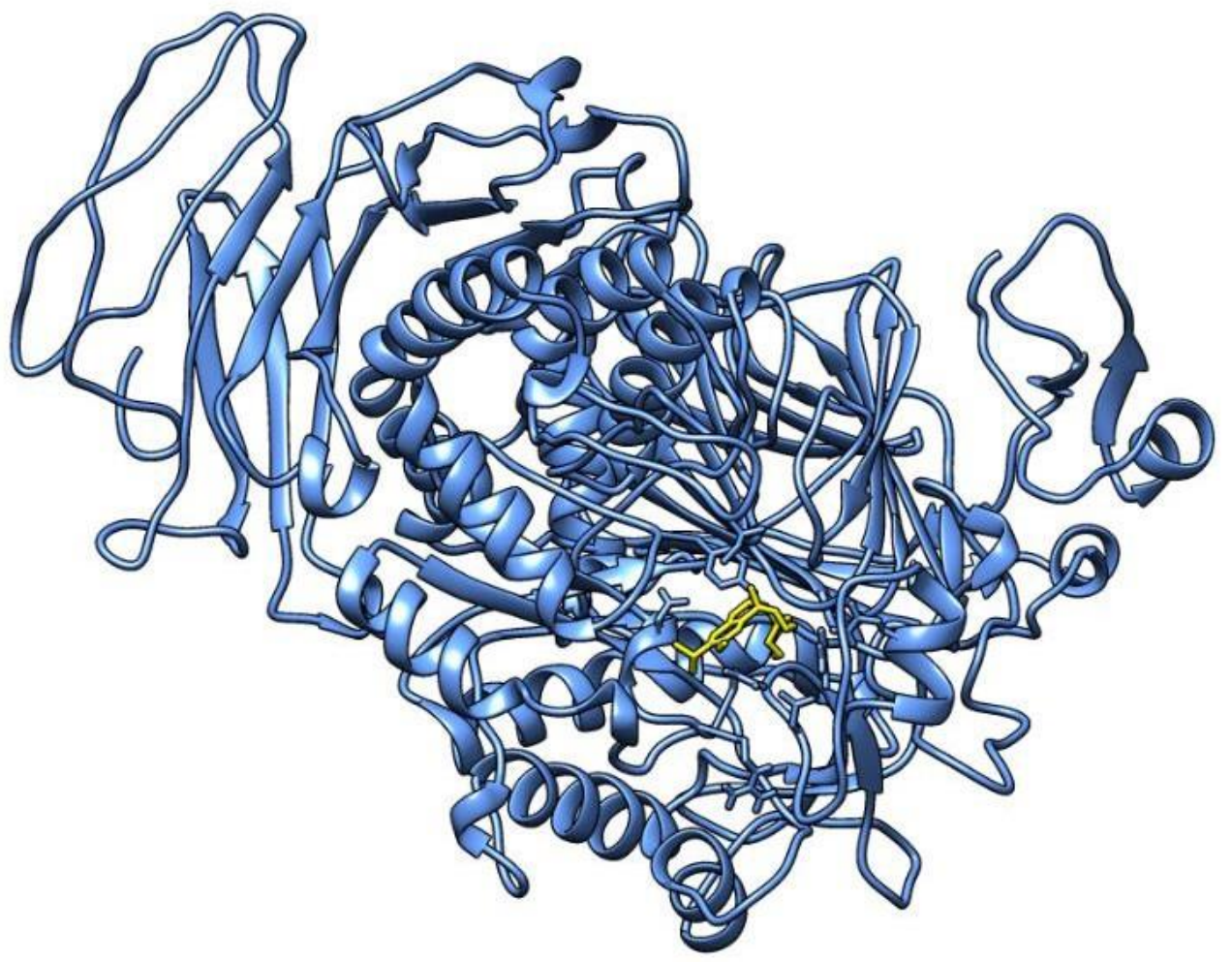

b.

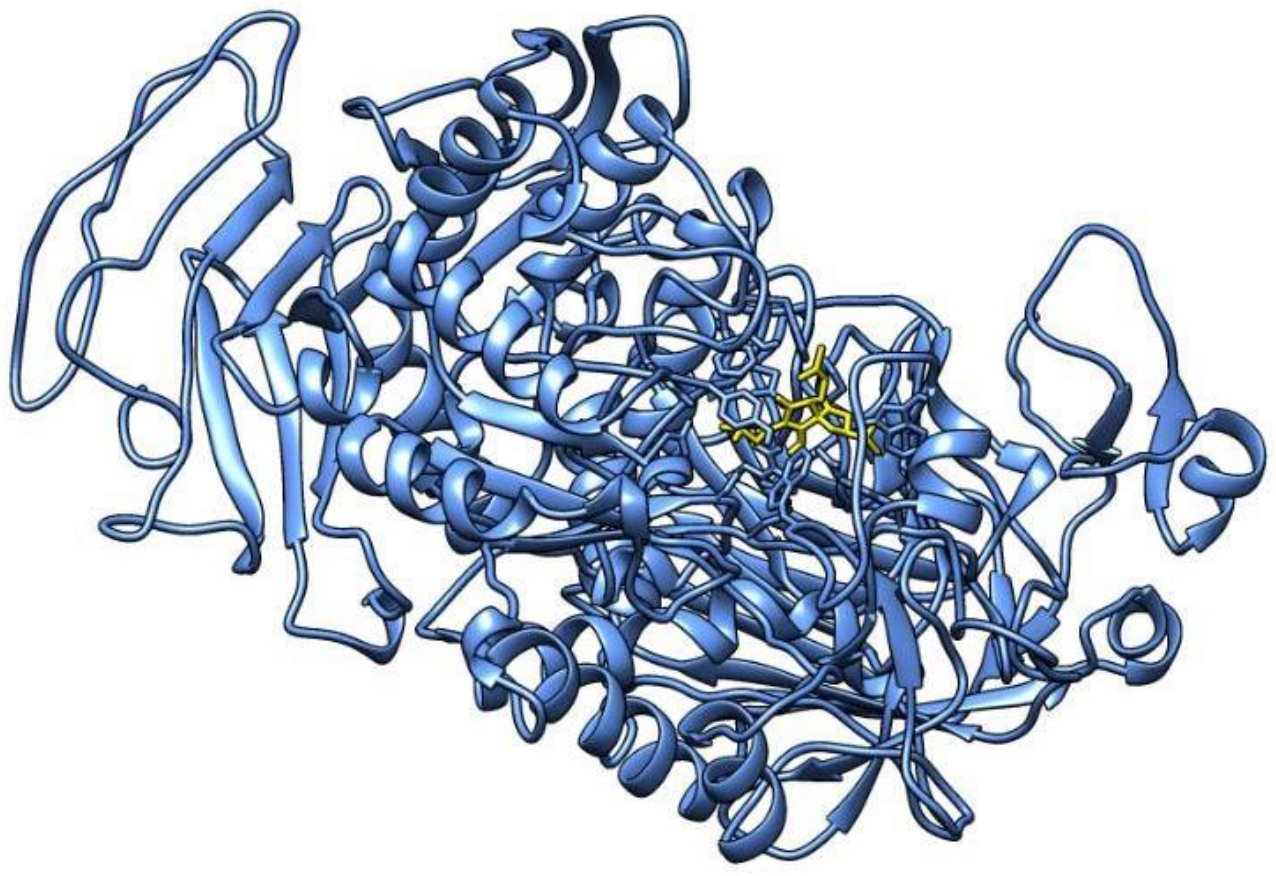

Fig. 1: GC/MS chromatograms for the analyzed tissues: a. Leaf; b. Stem; c. Root; d. Root Bark
Fig. 2: Distribution of volatiles in the analyzed tissues.
Fig. 3: Pair-wise similarity profile of the analyzed tissues.
Fig. 4: Redocking of acarbose to 3TOP (Human Maltase-Glucoamylase), native acarbose molecule depicted in blue while the redocked molecule is in red
Fig. 5: Leads docked to target protein a. CID – 240051 and b. CID – 533471, bound to Human Maltase-Glucoamylase enzyme (3TOP)